\documentclass[fleqn, usenatbib]{mnras}

\usepackage[T1]{fontenc}
\usepackage{newtxtext, newtxmath}

\usepackage{graphicx}	
\usepackage{amsmath}	
\usepackage{float} 
\usepackage{multirow}

\usepackage{fancyhdr}
\title[From customer survey feedback to software improvements]{From customer survey feedback to software improvements: Leveraging the full potential of data}

\author[Bertram et al.]{
Erik Bertram,$^{1,2}$\thanks{E-mail: erik.bertram@sap.com}
Nina Hollender,$^{1}$
Sebastian Juhl,$^{1,3}$
Sandra Loop,$^{1}$ and
Martin Schrepp$^{1}$
\\
$^{1}$SAP Deutschland SE \& Co. KG, Hasso-Plattner-Ring 7, 69190 Walldorf, Germany\\
$^{2}$Hochschule Fresenius Heidelberg, Sickingenstra\ss{}e 63-65, 69126 Heidelberg, Germany\\
$^{3}$University of Missouri, 230 Jesse Hall, Columbia, MO 65211, USA\\
}

\date{Version: \today}

\pubyear{2023}

\begin{document}
\label{firstpage}
\pagerange{\pageref{firstpage}--\pageref{lastpage}}
\maketitle

\begin{abstract}
Converting customer survey feedback data into usable insights has always been a great challenge for large software enterprises. Despite the improvements on this field, a major obstacle often remains when drawing the right conclusions out of the data and channeling them into the software development process. In this paper we present a practical end-to-end approach of how to extract useful information out of a data set and leverage the information to drive change. We describe how to choose the right metrics to measure, gather appropriate feedback from customer end-users, analyze the data by leveraging methods from inferential statistics, make the data transparent, and finally drive change with the results. Furthermore, we present an example of a UX prototype dashboard that can be used to communicate the analyses to stakeholders within the company.
\end{abstract}


\renewcommand{\headrulewidth}{0pt}
\pagestyle{fancy}
\fancyhead{}
\fancyhead[LE]{\Large \thepage\hspace{.5cm} \textit{E. Bertram et al.}}
\fancyhead[RO]{\Large \textit{From customer survey feedback to software improvements} \hspace{.5cm} \thepage}
\fancyfoot{}

\thispagestyle{empty}

\section{Introduction}
``What gets measured gets managed'' is a frequently cited quote that, however, is incorrectly attributed to the famous Austrian American management consultant Peter Drucker.\footnote{The quote originated from the columnist Simon Caulkin who summarized an argument in a paper by \citet{Ridgway1956}.} While this quote originated from a direct critique of the narrow and lopsided utilization of performance measures in a business context, the importance of quantitative metrics for companies, especially large and globally acting software enterprises, is growing \citep[see e.g.,][]{bauer2004,morgan2006,van2016}.

As such, one of their main targets is to increase the adoption of cloud services \citep{shimba2010} and to keep a stable customer base to drive revenue. To do so, every software company is eager to continuously improve the experience of their cloud product portfolio and deliver the world’s best software to customers. Furthermore, in today’s competitive markets, user experience (UX) quality is a pre-requisite for the long-term success of products \citep[see e.g.,][]{sward2007,ross2014}. Over time, competing products become more and more similar in terms of their functionality. For products with great UX, the experience becomes part of their brand, and therefore it is important to constantly measure the UX quality of products.

Furthermore, previous studies have revealed the role of change management and its impact on business processes and employees \citep[see e.g.,][]{mento2002, by2005, kettinger1995, hiatt2003, schollhorn2019}. However, companies face some notable challenges. For example, channeling end-users feedback into the software development process in an agile manner remains a big problem. Furthermore, it is often unclear how the gap between survey feedback and business insights can be overcome, and how these insights can be leveraged for continuous code improvements.

In this article, we aim to present a general and scalable end-to-end process and describe how companies can gather high-quality product feedback from end-users, analyze existing data with statistical methods, draw reliable conclusions, and finally feed this information into the development process. We will only use simulated data to illustrate our approach, as well as dummy products that we name Product A to Product F, for simplicity. All surveys presented below are publicly available on the internet and a common standard in many software enterprises \citep[see e.g.,][]{laugwitz2006, schrepp2019, fisher2019, lewis2013}. The article itself targets decision makers as well as developers, designers, and managers.

In general, we believe that establishing a smooth end-to-end process always requires a similar strategy, which we would like to sketch out on the following pages. The readers might feel free to adapt one or the other idea for their own purpose as well.

\section{Listening to customer end-users}
In the following, we describe several requirements and challenges that a company might face when listening to customer end-users with the help of surveys, including how to decide on appropriate survey items.

\subsection{Using standardized questionnaires to measure UX}
If we want to measure the user experience of a product quantitatively, we need to develop a clear understanding of what this term means semantically. The ISO 9241-210 norm \citep{isonorm} defines UX as a ``[…] person's perceptions and responses that result from the use or anticipated use of a product, system, or service.'' This definition entails two important aspects.

First, UX is a subjective perception of users concerning a product. Therefore, we must ask users, and different users might express different opinions. Such differences can, for example, result from personal preferences or diverse levels of expertise with a product. Thus, it is important to collect feedback from larger user samples to adequately cover the spectrum of different opinions.

Second, UX covers a wide range of different impressions. It does not only include quality aspects associated to working on typical tasks with the product (e.g., efficiency, controllability, or learnability), but also aspects like fun of use, or the aesthetic appeal of the user interface \citep[see e.g.,][]{fb_book_2019, hinderks_et_al_2021}.

For many years now, the UX discipline has developed several techniques to collect feedback from users. However, only a few options exist to measure the UX quantitatively. Online surveys are a very efficient tool to do so, since they allow collecting feedback from large user groups with low effort and costs. To get high-quality data, one needs to ensure that the measurement method fulfills common quality criteria, such as, e.g., objectivity, reliability, and validity. However, this is difficult to achieve if the questions are defined to calculate a UX score from the answers ad hoc. Nevertheless, there are many standardized UX questionnaires available that are carefully constructed, validated, and described in scientific publications \citep[see, e.g.,][]{fb_book_2019}.

\subsection{Choosing the correct UX questionnaires}
As explained above, UX is a heterogeneous concept and there are many questionnaires available that measure various aspects of UX. Hence, the first challenge is to decide which questionnaires to use. In general, we should measure those metrics that are most important for our users and that relate to common business goals \citep{Mc_Kinsey_2021, sauro_2016}. In addition, it is important to limit the number of questions as much as possible to reduce the drop-out rate, since long surveys can cause frustration on the end-user side. As an example, we will discuss the selection of metrics to evaluate a product portfolio below.

\begin{figure}
\centering
\includegraphics[scale = .45]{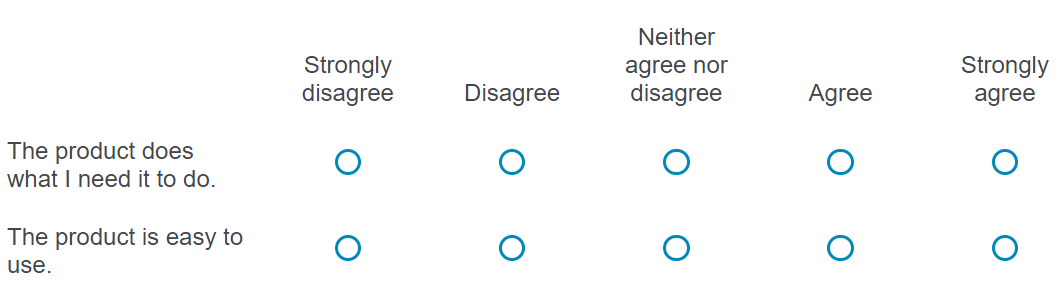}
\caption{The two questions of the UX-LITE questionnaire range on a five-point ordinal scale and measure usability and usefulness of a product.}
\label{fig:fig_1}
\end{figure}

Business applications are used for professional work. Typical users include, for example, developers of cloud applications, administrators, analysts, decision-makers, sales representatives, or accountants. For those users, both usability and usefulness are of high importance, which is why we need to ensure that those metrics are measured adequately. Therefore, the UX-LITE questionnaire is a natural choice, which consists of two questions, as shown in Figure \ref{fig:fig_1}.

The UX-Lite \citep{lewis_sauro_2021, finstadt_2010} realizes a concept similar to the Technology Acceptance Model \citep{davis_1989}, which assumes that user acceptance of a new technology is based on its perceived usefulness (first item of the UX-Lite) and perceived ease of use (second item of the UX-Lite). The answers are scored by numbers from 0 (Strongly disagree) to 4 (Strongly agree), while the sum of the two questions is calculated per participant. Thus, we get a rating between 0 (worst impression) and 8 (best impression). This score is then transferred to a range between 0 and 100 by multiplying it with a factor of $100/8 = 12.5$. By averaging over all participants we finally get a UX-Lite score for the product, which can be compared to the SUS score (also ranging between 0 and 100) and that allows us to use the well-established SUS benchmark \citep{lewis_sauro_2018, law_lewis_sumak_2020}. Hence, observed UX-LITE scores can be classified into eleven categories shown in Table \ref{tab:ux_benchmark}.

\begin{table}
\centering
\begin{tabular}{| l | c | c | c |}
Category & \multicolumn{2}{| c |}{Score Interval} & Percentile\\
\hline
A+ & 84.1 & 100 & 96 - 100\\
A & 80.8 & 84.0 & 90 - 95\\
A- & 78.8 & 80.7 & 85 - 89\\
B+ & 77.2 & 78.8 & 80 - 84\\
B & 74.1 & 77.1 & 70 - 79\\
B- & 72.6 & 74.0 & 65 - 69\\
C+ & 71.1 & 72.5 & 60 - 64\\
C & 65.0 & 71.0 & 41 - 59\\
C- & 62.7 & 64.9 & 35 - 40\\
D & 51.7 & 62.6 & 15 - 34\\
F & 0.0 & 51.6 & 0 - 14
\end{tabular}
\caption{Benchmark categories used for the UX-Lite, including score intervals and percentiles. Please note that American school grades are used to classify the categories.}
\label{tab:ux_benchmark}
\end{table}

For example, a value of 82 would indicate that the product can be assigned to category A (i.e., around 90\% of the products in the benchmark data set show a lower and 5\% a higher score while the remaining 5\% of products fall in the same category). Such a value indicates a pretty good quality, whereas if a value of 45 is measured, the product would lie in category F (i.e., it is amongst the 14\% of products in the benchmark data set that show the lowest scores), indicating a bad quality.

\begin{figure}
\centering
\includegraphics[scale = .45]{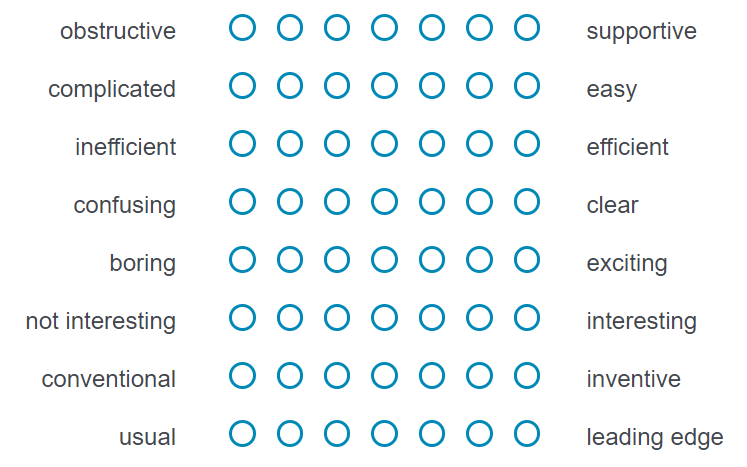}
\caption{The UEQ-S is a short survey with eight semantic differential questions, raging on a seven-point ordinal scale.}
\label{fig:fig_2}
\end{figure}

To keep users engaged and motivated, it is also important that they perceive a product as interesting and enjoyable. There are few surveys that measure joy of use. We selected the UEQ-S, a short version of the User Experience Questionnaire \citep{laugwitz2008, schrepp_2017}, which is a short survey with eight semantic differential questions, shown in Figure \ref{fig:fig_2}. All eight items are grouped into two subscales. The first four items measure usability of the product, the last four items fun-of-use or interest.

Items are scored from $-3$ to $+3$ from left to right. The mean for the two subscales is simply the mean score over all items and participants in a scale. The overall score is the mean value over all eight items.

The UEQ-S provides also a benchmark (see Table \ref{tab:ueq_benchmark}) that classifies scores measured for the subscales into the five categories \textit{Excellent}, \textit{Good}, \textit{Above Average}, \textit{Below Average} and \textit{Bad}. The logic behind this benchmark is similar to the benchmark used for the UX-LITE (i.e., an extensive reference data set of more than 450 studies was split into these five groups).

\begin{table}
\centering
\begin{tabular}{| p{1cm} | c | c | c | c |}
Category & User Experience & Usability & Fun of Use & Percentile\\
\hline
Excellent & 5.58 - 7.00 & 5.74 - 7.00 & 5.59 - 7.00 & 91 - 100 \\
Good & 5.31 - 5.57 & 5.55 - 5.73 & 5.20 - 5.58 & 76 - 90 \\
Above Average & \multirow{2}{*}{4.98 - 5.30} & \multirow{2}{*}{5.17 - 5.55} & \multirow{2}{*}{4.85 - 5.19} & \multirow{2}{*}{51 - 75} \\
Below Average & \multirow{2}{*}{4.59 - 4.97} & \multirow{2}{*}{4.72 - 5.16} & \multirow{2}{*}{4.35 - 4.84} & \multirow{2}{*}{26 - 50} \\
Bad & 1.00 - 4.58 & 1.00 - 4.71 & 1.00 - 4.35 & 0 - 25 \\
\end{tabular}
\caption{The UEQ-S benchmark consists of five categories (Excellent, Good, Above Average, Below Average, Bad). Shown are the scores for user experience, usability, fun of use, and the corresponding percentiles.}
\label{tab:ueq_benchmark}
\end{table}

\begin{figure}
\centering
\includegraphics[scale = .45]{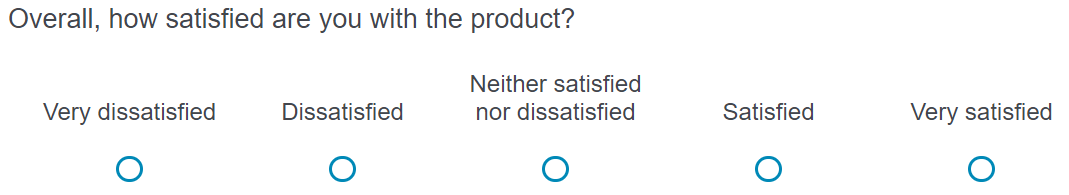}
\caption{The PSAT score accounts for customer satisfaction and also ranges on a five-point ordinal scale.}
\label{fig:fig_3}
\end{figure}

Clearly, the UX is not the only driver that affects the success of a product. Instead, both the support and documentation quality or the availability of the application in the cloud are crucial aspects as well. To account for those, we added the PSAT (Product Satisfaction)\footnote{Note that in some instances, the PSAT score is referred to as 'Customer Satisfaction' score. For consistency reasons, we use the term PSAT throughout this publication.} score, which is calculated from one single question, as indicated in Figure \ref{fig:fig_3}.

The PSAT score is calculated as the percentage of participants that answered with \textit{Very satisfied} or \textit{Satisfied}, thus ranging from 0 (worst) to 100 (best). It is an easy to interpret metric that represents the overall satisfaction of a product. The American Customer Satisfaction Index (https://www.theacsi.org/) is based on the PSAT and a yearly benchmark is published for different industry sectors, including software products. 

\begin{figure}
\centering
\includegraphics[scale = .45]{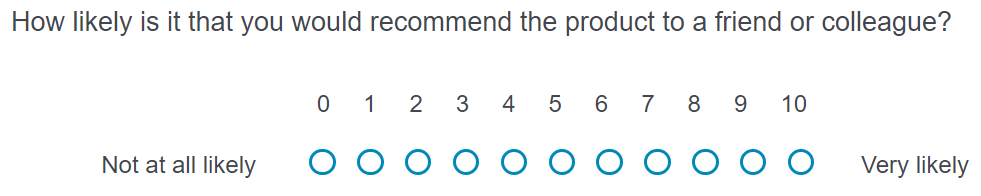}
\caption{The NPS ranges on a simple ten-point ordinal scale (from not at all likely to very likely) and measures customer satisfaction.}
\label{fig:fig_4}
\end{figure}

For cloud products, a high renewal contract rate is crucial for generating revenue and achieving long-standing business goals. Thus, the loyalty of users or customers is of vital importance. The NPS (Net Promotor Score) is a widely used method to measure customer loyalty and will therefore be included as well. It is calculated from the single question shown in Figure \ref{fig:fig_4}.

Responses are coded as \textit{Promoters} (with scores of nine or ten), \textit{Passives} (with scores of seven or eight) and \textit{Detractors} (with scores from zero to six). The NPS score is calculated by subtracting the share of \textit{Detractors} from the share of \textit{Promoters}. The score ranges from $-100$ (worst possible) to $+100$ (best possible). The core idea of the NPS \citep{reichheld} is that people that are very positively impressed by a product or service (\textit{Promoters}) will recommend this product or service to others and thus increase the user base.

These four metrics seem to cover the most central aspects for a portfolio of cloud applications. However, depending on the type of applications and the specific business goals, other combinations of metrics might be required. Thus, one must carefully investigate the importance of different requirements and select the proper metrics accordingly. 

A big advantage of using established survey items in standardized questionnaires is the availability of benchmarks, which consist of a larger set of product evaluations that use the corresponding questionnaire. Hence, it is possible to compare own results with the results of the products in the benchmark data set, helping to interpret and contextualize the numbers.

\subsection{Adding supplementary information}
Metrics like those described above provide information about the overall UX quality. However, the plain numbers do not explain which product features must be improved to obtain better results in the future. To collect information that helps to interpret the quantitative results, we recommend including two open comment fields in the survey. They ask participants to comment on the strong (\textit{What did you like about the product?}) and weak aspects of a product (\textit{What should be improved?}). Different users may have different opinions concerning some product features. Thus, it is not unusual to receive both positive and negative comments concerning the same product feature. Therefore it is important to always ask for positive and negative feedback to avoid a biased impression. Analyzing these comments helps to understand the ratings and is crucial to transfer the interpretations into concrete actions.

Utilizing the capabilities of generative AI tools and Large Language Models like GTP-4 it is possible to automatically summarize and categorize a vast amount of user comments. This can save time and resources while allowing the extraction of valuable information from unstructured text data.

In addition, questions about the product usage can be captured (e.g., role of the user, age, experience with the product, frequency of use), which provide further insights into a product portfolio. For example, one could imagine a development team that improved a workflow for a specific user role in their product. While the overall score across all user roles might not have improved notably, looking at the scores for the target user role might show an improvement of the score some time after the release of the feature.

\subsection{Single vs. multiple surveys}
In large software enterprises with a diverse product landscape made of different UX maturity levels, additional challenges might arise. For example, some product teams might already have started to use surveys to collect end-user feedback, and these surveys might use different UX metrics than those described above. In such cases, teams might refuse changing to the new set of metrics, and in fact, the availability of data collected in the past is a valuable source of information since it offers a baseline against which the measurement of new versions could be compared. Furthermore, it is important to choose common metrics to be able to compare the UX quality of all products with each other. To compromise, one could enhance the existing surveys with the most important new metrics, while it must be ensured that the new survey does not get too long. In our case, we decided to add at least the two UX-Lite questions and the PSAT question to all existing surveys to achieve some comparability.

In addition, special products may want to ask special product-related questions to gain deeper insights. Such questions may be necessary only for a certain point in time. For example, if a new feature is delivered it might be helpful to capture the user satisfaction containing this feature with a dedicated question in the survey. If some conclusion could be reached from the answers, this question can be removed again to shorten the survey. Thus, a successful approach requires some openness for application specific needs. Technically, this is often a challenge since a substantial number of different surveys might be required that need to be consolidated for analyzing the common metrics.

\subsection{Coordinating a cross-product initiative}
Depending on the size of the organization, one may have or need smaller groups to organize such a large-scale measurement project. Having a contact person, typically a UX designer, a product manager, or a small research operations team for the specific product area can help to channel the survey into the respective development process. Having contacts for each product is helpful when change is required, e.g., when you want to support auto-triggered in addition to user-triggered surveys. If code changes are required, the contact champions can then talk to their product teams about it. However, note that even with the champions in place it can be a huge coordination effort to involve multiple products. Keeping a record of who agreed to what and when is crucial for tracking purposes of the initiative.

\section{Gathering end-user data}
After deciding upon the right metrics to use, applying them to customer end-user via dedicated surveys is a vital step in the whole measurement process. In the following, we will sketch several facts that one should consider during the data collection phase.

\subsection{Data protection regulations}
Before reaching out to users via online surveys, we advise to carefully check existing data protection regulations. Although the feedback is provided anonymously, capturing personal data in the realm of the General Data Protection Regulation (GDPR) can in most cases be avoided. However, one should ensure that specific information such as, e.g., IP addresses, demographic information, or the company name is not collected carelessly. A grey area certainly is an open text field, into which a user can enter personal information. It is advisable to add a disclaimer in the instruction of the survey or directly above the comment field that instructs users to avoid this. 

If one needs to collect data that is detailed enough to identify a person, you must add a special data privacy statement to the survey and ensure that the responses are stored only if the user agrees to this (especially in the European Union). In addition, you need to define a clear process that ensures that such personal information is deleted if it is no longer needed or after a certain amount of time.

If the application is used by employees of corporate customers, you should also check whether contractual agreements allows to send the survey to these employees.

\subsection{Ensuring an adequate feedback sample size}
A challenge that might occur in the data collection phase are weak feedback streams from the end-users, causing low-number statistics. Nevertheless, there exist different channels that can be used to ask for user feedback:

\begin{itemize}
\item Links or buttons directly in the UI of a product that open a feedback form or a longer survey when a user clicks on these elements. 
\item Automatic triggers that ask users for feedback while they work with the product. For example, such a trigger could launch a dialog after the user worked five minutes in the application. The dialog asks the user to provide feedback, and if the user agrees, a feedback form is opened. It is also possible to trigger the request for feedback when users complete a task or perform a special action. 
\item A link to a survey can be distributed via dedicated e-mail or social media channels. 
\end{itemize}

Each of these channels have advantages and disadvantages. For example, feedback forms that are accessible in the UI (either launched by a passive feedback button or an active automatic trigger) are easy to implement and reach the users when they are working with the product. Thus, the feedback is not provided in retrospective, but rather gives a direct impression of how the product is perceived by the customer. However, since users would like to work in the system rather than spend their time answering questions, they tend to not accept longer surveys in this situation (which is especially true for the case where feedback is requested by automatic triggers). Hence, such in-system feedback mechanisms should be short to avoid high dropout rates. In contrast, feedback requests distributed over e-mail or links via social media channels allow longer surveys with more detailed questions.

Especially automatic triggered requests for feedback are often perceived as disturbing or annoying. Thus, they negatively influence the UX and may even force users to provide bad ratings, since they are angry about the interruption.

Feedback mechanisms launched by a passive button or a link typically generate less feedback. They are clicked if users detect a problem or positive behavior of the user interface and want to report it to the development teams. Hence, there may be a bias in the responses towards very positive or very negative ratings. 

Some products may have a poor response rate, because they have fewer users or the users do not want to offer feedback. In this case it can be required to collect feedback over all available channels, for example, by an email campaign delivered by an account manager who knows the customer personally or by a well-known thought leader, a feedback button in the UI, or a request for feedback in social media channels. 

To address these issues, we recommend implementing different feedback mechanisms for different products. All products contain a feedback button that launches a short feedback form that contains the PSAT and the UMUX-Lite questions, and a comment field. In addition, e-mail campaigns can be used to collect more details for important products. The survey launched from such an e-mail campaign might contain questions about demographic variables, usage behavior (e.g., experience, frequency of use, etc.), the four questionnaires PSAT, UX-Lite, UEQ-S, and NPS together with two comment fields that ask for positive and negative feedback. 

To take advantage of the benefits of all different questions, you can also choose to have multiple surveys for a product, e.g., by using a short survey with standardized questions plus an open text field, where this survey is always available over a feedback button in the system, and a longer survey with the standardized questions plus some context question including product-specific questions, distributed via e-mail campaigns.

\subsection{Biases introduced by the different channels}
\begin{figure}
\centering
\includegraphics[scale = .4]{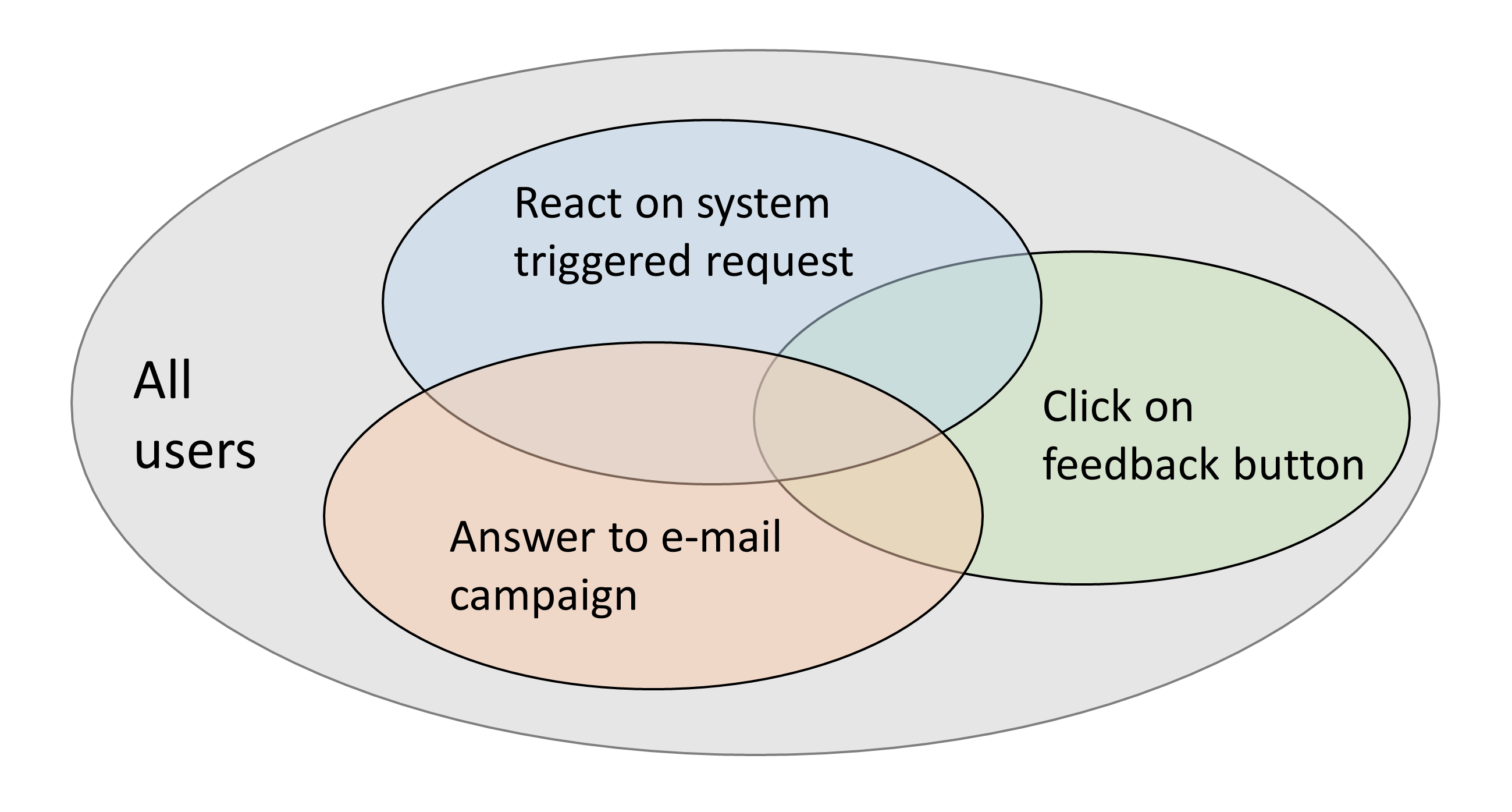}
\caption{Different biases introduced by our method to collect end-user feedback.}
\label{fig:biases}
\end{figure}

In general, one needs to be cautious when comparing UX metrics collected over different channels, since each method comes with an associated bias (see Figure \ref{fig:biases}). For example, let’s assume that we place a feedback button in the UI. Users that are unsatisfied or satisfied concerning the product will click on this button more frequently than users with a neutral opinion. If you actively ask for feedback, you may motivate some users that otherwise might not have clicked the button, while other users might feel disturbed by such requests, introducing yet another bias. Hence, any feedback collected via online surveys will always contain some bias, since none of the feedback channels will deliver a representative sample of all user perceptions.

If you want to compare ratings obtained by different methods, you should be careful as differences may be caused by the method used to collect the data rather than true differences in the KPI. However, we are often interested in the ratings of all end users, although in practice we only have survey responses from a small user sample. In this case, we want to take the sampled user information only and extrapolate the results to the general user population, which is the set of all end-users. Nevertheless, since different users have different opinions, the results we receive are likely to change if we ask a different sample of users.

\section{Applying statistical methods}
In the following, we describe how one can use well-known statistical methods from inferential statistics to gain a deeper understanding of the UX perceptions of the users.

\subsection{Analyzing the data}
As a first step, we use the survey data to calculate aggregated measures, such as mean values, standard errors, and proportions of the KPIs outlined above. These descriptive measures allow us to gain insights into how the users evaluate their experiences with the products. However, we are not only interested in how the respondents of our surveys evaluate the UX of our products. Instead, we would rather like to know how the overall user population is rating the UX. The mean values we calculate from the sample may not be representative of all users since it is possible that we only asked a very specific subset of users by chance. If we would have asked any other sample of users, our results might have looked different. Consequently, we need to account for the uncertainty caused by asking only a sample and not the entire user population.

\subsection{Learning about the general user population}
A simple example illustrates this problem. Consider a product with 1,000 end users in total. We would like to know how satisfied these users are on average with the product. If we would have asked all 1,000 users to rate their product satisfaction on a five-point scale, we would see that the true average is $\mu = 3$.

In reality, however, we do not have responses from all 1,000 end users, but only from a small sample of, e.g., 50 randomly selected users. Taking the responses from these 50 users, we might get an average satisfaction rating of $\bar{x}_1 = 2.76$, calculated by $\frac{1}{n}\sum x_i$, where $n$ is the total number of responses and $i \in n$ denotes an individual response. While this value is fairly close to the true mean of 3, we could have drawn a different sample of 50 users which would have yielded different results. In fact, taking a second sample from all 1,000 users and calculating the average product satisfaction provides a score of $\bar{x}_2 = 3.24$. In theory, we could have asked yet another sample of users and most likely received again a different average product satisfaction score.

\begin{figure}
\centering
\includegraphics[scale = 0.3]{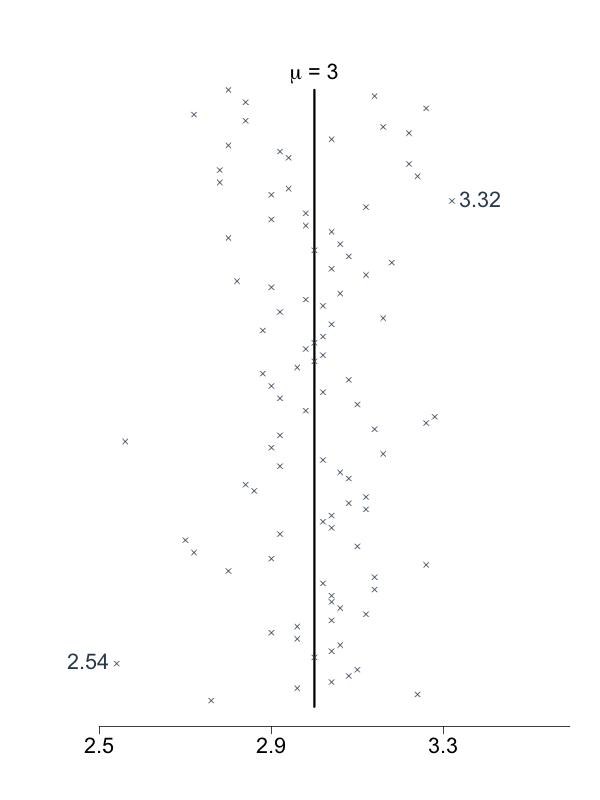}
\caption{Drawing 100 samples of 50 users from a user population with $\mu = 3$ and recording the respective sample means measuring $\bar{x}_1$ to $\bar{x}_{100}$. In this case, the sample means range from $\bar{x}_7 = 2.54$ to $\bar{x}_{82} = 3.32$.}
\label{fig:means}
\end{figure}

By repeating this process of randomly sampling 50 users from all 1,000 end users 100 times and recording the respective average satisfaction score, we would get the results as shown in Figure \ref{fig:means}. Clearly, our assessment of the end users‘ product satisfaction differs depending on the sample used. Each time we ask a different sample of users, we get slightly different results. While the values fluctuate around the true value of 3, we end up with values between 2.54 and 3.32 in some situations simply because we have asked a specific sample of users by pure chance.

To account for this, we utilize different methods from inferential statistics in order to use the responses from the sampled users to learn about the general user population. With these techniques, we can draw conclusions about the general user population and quantify the uncertainty caused by asking only a small subset of users.

\subsection{Confidence Intervals}
Since the sample mean $\bar{x}$ may not equal the population mean $\mu$, we calculate confidence intervals, indicating the range in which the true population mean $\mu$ lies with a pre-defined level of confidence. Using the sample mean $\bar{x}$ and the sample variance $s^2$, confidence intervals for the means are calculated by:

\begin{equation}\label{eq:CI}
\bar{x} \pm t(df, \alpha/2) \times \sqrt{\frac{s^2}{n}},
\end{equation}
where $n$ represents the number of survey responses and $t(df, \alpha/2)$ corresponds to the critical value derived from a two-tailed $t$-distribution, with $\alpha$ being the desired threshold for statistical significance and $df$ representing the degrees of freedom given by $df = n - 1$.\footnote{The $t$-distribution is a generalization of the normal distribution. While it is also a continuous and symmetric probability distribution, it has more mass at its tails if $n$ is finite. This makes the results of statistical tests more conservative as the critical value a test statistic needs to reach to be statistically significant is higher than the critical value derived from the standard normal distribution. As $n \rightarrow \infty$, the $t$-distribution approaches the normal distribution.} Since we stick to the convention of reporting $95\%$ confidence intervals, we use $\alpha = 0.05$. The sample variance is given by:

\begin{equation}\label{eq:SD}
s^2 = \frac{1}{n-1} \sum_{i=1}^n(x_i - \bar{x})^2.
\end{equation}

Equation \ref{eq:CI} defines the lower and upper bound of the $95\%$ confidence interval which is symmetric around the point estimate $\bar{x}$ by design. The formula also shows that the widths of confidence intervals is determined by two factors: the number of responses $n$ and the variance $s^2$. The more responses we have, the narrower the confidence interval and the more precise our estimate of $\mu$. Conversely, the larger the variability in survey responses, the wider the confidence interval.

Constructing confidence intervals around proportions (e.g., for the PSAT score) requires a slightly adjusted formula. Specifically, we use a standard normal distribution rather than the $t$-distribution to derive the critical value and replace the sample standard deviation $s^2$ by $\hat{p}(1 - \hat{p})$, where $\hat{p}$ is the estimated proportion from the sample:

\begin{equation}\label{eq:CIprop}
\hat{p} \pm z_{\alpha/2} \times \sqrt{\frac{\hat{p}(1 - \hat{p})}{n}}.
\end{equation}

Again, the more responses we receive, the narrower the confidence intervals. Furthermore, confidence intervals become wider if $\hat{p}$ is either close to $0$ or $1$.

While confidence intervals provide important information on the range in which the true value $\mu$ lies (given a specific level of confidence) if we would have asked all users, many business use cases require the comparison of KPIs across products or over time. For example, we may be interested in knowing whether the PSAT score of a product increases from one quarter to the next. Alternatively, we may also be interested in knowing whether users rate the usefulness of one product higher as compared to another product.

\subsection{Hypothesis Testing}
Confidence intervals have only limited utility in answering these questions. The reason is that the overlap in confidence intervals is not sufficient to gauge whether or not differences are caused by chance alone. Even if confidence intervals do overlap, their difference may still be statistically significant \citep[see, e.g.,][]{schenker2001}.

To overcome this limitation, we utilize additional techniques from inferential statistics for hypothesis testing. Specifically, we want to know whether a KPI measured for two groups differ in the user population by asking a small subset of users. Denoting the true population parameters (the value we would obtain if we would have asked all users) of a KPI by $\mu_1$ and $\mu_2$, we test the null hypothesis of no difference in means. The corresponding alternative hypothesis is undirected and states that both population parameters differ:

\[
\begin{array}{rl}
H_0: & \mu_1 = \mu_2\\
H_1: & \mu_1 \neq \mu_2
\end{array}
\]

For the difference in means, we perform a two-tailed independent samples Welch's $t$-test to evaluate the null hypothesis. In contrast to the Student's $t$-test, this method has a superior performance if both groups have unequal variances and their sample sizes differ as it does not use a pooled variance estimate but rather a combination of the group variances \citep{welch1947}.

We first use the groups' sample means $\bar{x}_1$ and $\bar{x}_2$ and their sample variances $s^2_1$ and $s^2_2$ as shown in Equation \ref{eq:SD} to calculate the test statistic:

\begin{equation}\label{eq:teststat}
t = \frac{\bar{x}_1 - \bar{x}_2}{\sqrt{\frac{s^2_{x_1}}{n_1} + \frac{s^2_{x_2}}{n_2}}}.
\end{equation}

Equation \ref{eq:teststat} illustrates that, all else equal, $t$ becomes larger as i) the difference in observed means increases, ii) the variances in the samples decreases, and iii) the number of responses per group increases. 

Next, we determine the critical value the test statistic $t$ needs to exceed in order to be statistically significant. To this end, we again set the threshold for statistical significance to $\alpha = 0.05$ and approximate the degrees of freedom by the Welch-Satterthwaite equation \citep[e.g.,][]{Satterthwaite1946, Satterthwaite1941}:

\begin{equation}
df = \frac{\left(\frac{s^2_{x_1}}{n_1} + \frac{s^2_{x_2}}{n_2}\right)^2}{\left(\frac{s^4_{x_1}}{n_1^2 (n_1 - 1)} + \frac{s^4_{x_2}}{n_2^2 (n_2 - 1)} \right)}.
\end{equation}

Using $\alpha$ and $df$, we derive the critical value $t(df, \alpha/2)$ from a $t$-distribution. If $t > t(df, \alpha/2)$, the difference we find in the sample is statistically significant. Consequently, we can reject $H_0$ and conclude that the true population parameters $\mu_1$ and $\mu_2$ differ based on the samples we used and the pre-defined confidence level.

Conversely, if $t \leq t(df, \alpha/2)$, we cannot conclude that the difference we find between $\bar{x}_1$ and $\bar{x}_2$ also exists in the user population. That is, there might be a true difference in $\mu_1$ and $\mu_2$ but given the data we have, we cannot be confident enough that the difference we find in the sample is also present in the overall user population.

To test the statistical significance of a difference in proportions (e.g., for the PSAT score), we follow the same basic steps. However, instead of Equation \ref{eq:teststat}, we use the observed proportions from the samples $\hat{p}_1$ and $\hat{p}_1$ and the following formula to derive the test statistic:

\begin{equation}
z = \frac{\hat{p}_1 - \hat{p}_2}{\sqrt{\frac{\hat{p}_1(1-\hat{p}_1)}{n_1} + \frac{\hat{p}_2(1-\hat{p}_2)}{n_2}}}.
\end{equation}

Furthermore, instead of using a $t$-distribution, we derive the critical value $z_{\alpha/2}$ from a standard normal distribution using the same threshold for statistical significance of $\alpha = 0.05$.

Again, if $z > z_{\alpha/2}$, the difference in the proportions is statistically significant and our analysis indicates that there is a true difference in the user population. If $z \leq z_{\alpha/2}$, we cannot rule out the possibility that any difference we find in the samples are caused by chance alone and that there is no difference in the user population.

While identifying statistically significant changes over time or across products is crucial for data-driven decision-making in a business context, it is important to note that statistical significance does not imply substantive significance. A very minor change in a KPI might be statistically significant (e.g., because the sample size is huge or there is little variation among survey responses), but of such a small magnitude that it is substantively negligible. These statistical hypothesis tests provide pivotal information in order to avoid basing decisions on random fluctuations of sampled data but they should not be viewed in isolation from the relevant business context.

\section{Making the data and the analyses transparent}

Typically, the UX of a product cannot be improved by a single person alone. Instead, the whole team needs to be accountable to increase the product satisfaction. Thus, we recommend that everyone in a company gets access to the data and the analyses to help influence the UX, from executives to individual contributors.

\subsection{Data storytelling using a dashboard and a report}
To achieve this goal, one could create a dashboard (e.g., by using a tool like SAP Analytics Cloud) that displays all the KPIs and automatically creates the relevant statistical measures described above. In addition, one could also create a UX measurement report and make it accessible for everybody in the company. A dashboard can be great for deeper analysis to look for areas to research, and if it supports filtering, showing historical trends of the KPIs or context information gathered in the surveys it can provide huge value for employees.

\begin{figure*}
\centering
\includegraphics[scale = .39]{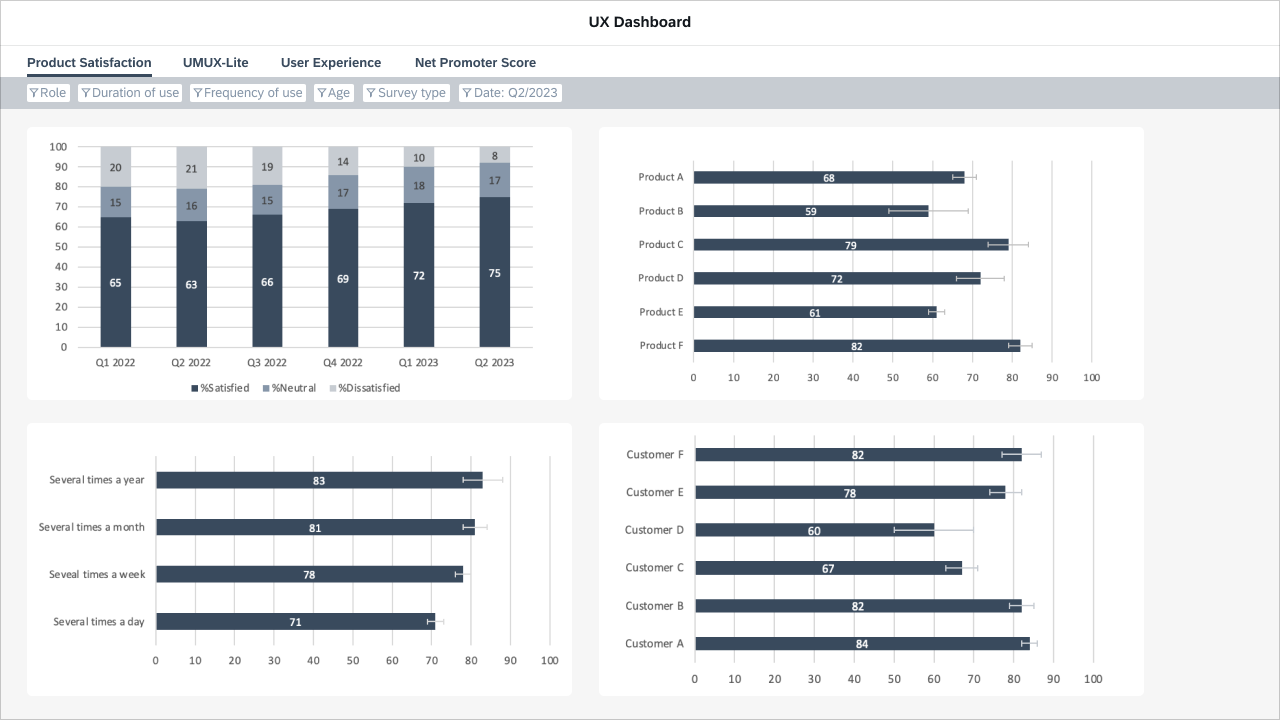}
\caption{This UX prototype dashboard is an example screen showing the product satisfaction scores and their $95\%$ confidence intervals for six products over time. The charts at the bottom show the scores sorted by customer and frequency of use. By combining different filters at the top one can try to get more information out of the survey data and analyze different use cases.}
\label{fig:dashboard_psat}
\end{figure*}

\begin{figure*}
\centering
\includegraphics[scale = .39]{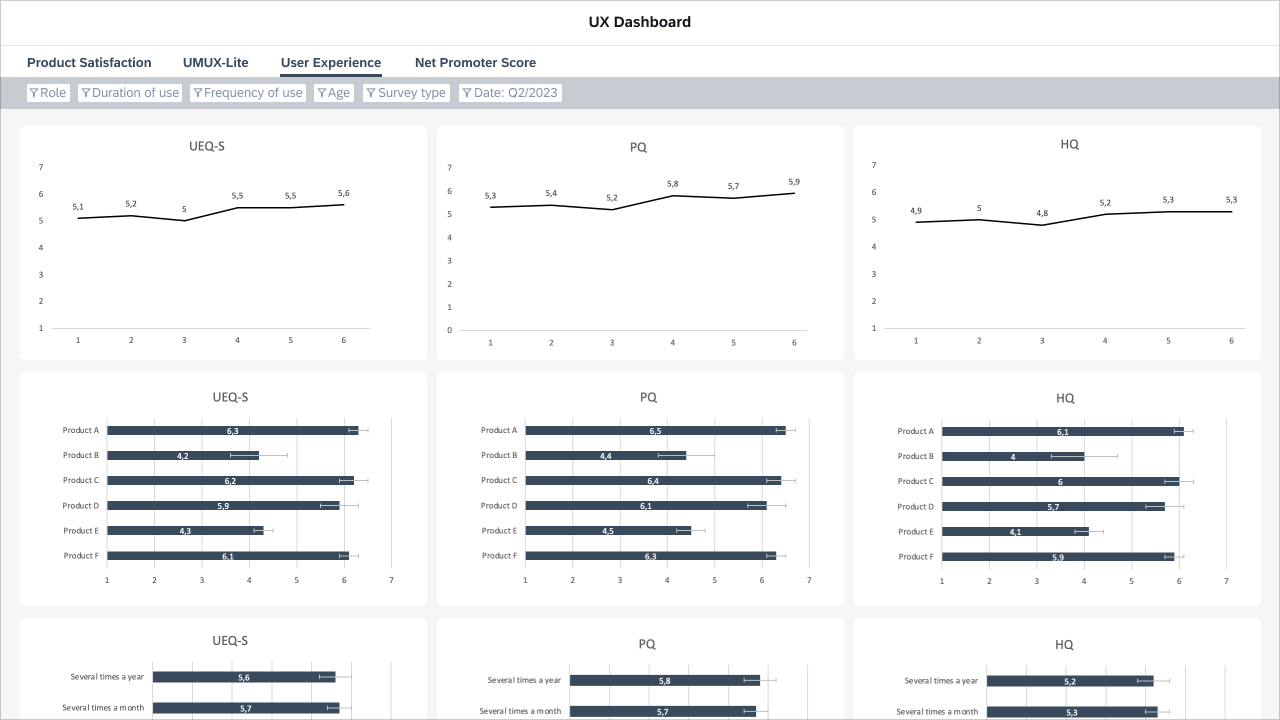}
\caption{Another UX prototype dashboard showing the UEQ scores and their $95\%$ confidence intervals for six products over time. It also shows the split between the products' pragmatic and hedonic qualities. Again, by combining different filter settings at the top one can try to extract additional information out of the data set.}
\label{fig:dashboard_ueq}
\end{figure*}

Figure \ref{fig:dashboard_psat} and Figure \ref{fig:dashboard_ueq} illustrate how a UX dashboard can look like in practice, using imaginary product names and mock data. While Figure \ref{fig:dashboard_psat} showcases an example screen for the PSAT score, Figure \ref{fig:dashboard_ueq} provides information on the UEQ score. The example UX dashboard shown here has a header with the option to switch between different KPIs. Business users interested in, e.g., the Net Promoter Score can easily navigate to this KPI by clicking on the respective button in the header and find more detailed information there. In the product satisfaction screen, the stacked bar chart displays the distribution of users grouped by their level of satisfaction across time. If the PSAT score drops, for example, this chart allows users to see whether previously satisfied users become neutral or rather dissatisfied users. The other three charts provide information on the share of satisfied users across different levels of aggregation. The example shows that users who frequently interact with the products are less satisfied than users who use the products just occasionally.

For the UEQ example screen shown in Figure \ref{fig:dashboard_ueq}, the evolution of some scores over time is shown as three line charts, allowing the users to look at the overall UX ratings as well as the users' evaluations of the products' pragmatic and hedonic qualities. Again, the example screen also features a product-level split alongside the split by frequency of use.

In addition to the information provided by the different charts, there are several filtering options available in the header that allow the user to take a closer look at a specific subset of the data. For example, consider the scenario where you are working on a product, while you want to learn more about its perception on the customer side. In the dashboard, you could filter for this specific product and recognize that the customer satisfaction is down since last quarter, but upon further inspection, you observe that customer satisfaction is simply recovering from a previous spike in the score. You could slice-and-dice to see what specific user roles say about the product. For example, you might find that administrators like the product, while DevOps colleagues are not as enthusiastic. Additionally, you might find an interesting pattern where people who are new to the product are not happy, yet the longer people have been using it, the more they like it. This could be explained as new users struggle to learn the product and how to make use of it, but after they mastered it, their satisfaction increases. You might also see that one company contributed many responses, and all scores from this company lie below the average value. This could happen when a company encourages their users to respond to the survey to have their thoughts heard. Reading through the free text-field comments can help you to confirm this. Such an analysis could trigger many ideas for further research.

Moreover, a report can be helpful as well for those who just requested a status update or snapshot of what the users think of the products. A report is generally easy to consume, particularly if it draws attention to important information. For instance, the report could highlight significant changes since the previous report. It could also provide a product summary page with the KPIs for each product including a sampling of the comments from the surveys. One may decide whether to host the reports in an internal broadly accessible location or to also send out the report once released to a distribution list. 

\subsection{Fostering adoption through appropriate data visualization}
To ensure that the stakeholders take away the appropriate insights, the dashboard and reports must be easy to consume. The International Business Communications Standards \citep{IBCS}, also called IBCS, emphasizes standard charting notation and minimizes unnecessary colors except when semantically relevant to help your reports and dashboards being easier to understand. If you are using SAP Analytics Cloud, you could also follow the 10 golden rules of dashboard design as described in \citet{DashboardDesignWithSAC}. Furthermore, one needs to ensure that the dashboard does not suffer major performance issues, which might lead to a low adoption of the application \citep{bertram2024}.

To guarantee that many product teams use these resources to improve the products, the adoption needs to be high. Change management theory suggests that you should involve the stakeholders early so that you can hear their concerns and pivot appropriately \citep{yeoh2008}. You could take early designs of the dashboard to the product teams and request a usability testing of the dashboard to hear about the concerns, and if you hear that some UX teams fear that they will be judged based on the outcome, you might convey this concern to the executive sponsors to ensure that they indicate that no teams will be judged, and that the entire product team is required to make improvements. In addition, you could also organize onboarding sessions explaining how to slice and dice the data and how to interpret the statistics. 

Another concern might be that the product adoption could drop when some UX improvements will not result in immediate score improvements. In this case, one could include in the roll-out messaging that people should trust that the UX is important and that eventually the scores will rise with enough effort and iterations.

To ensure all members of the product teams are aware of the dashboard and reports, the exact location of these resources should be widely evangelized, e.g., by referring to them in portals, meetings, internal conference, newsletters, etc. Clearly, they should be accessible to everyone without additional overhead of granting access rights.

\section{Driving change with insights}
Ultimately, the UX KPIs are measured to drive improvements in the products. Even though these are UX KPIs, the UX team alone should neither be seen as the enforcer, nor as being solely accountable for any product improvements.

To achieve the necessary level of product experience improvements, there are more steps along the way, while gathering UX KPIs are merely the first step. The product teams should set target KPI values, and periodically review the current state. If the targets are not reached, the product teams should decide on what actions to take. Gathering survey responses typically will not indicate how to change the product, but merely points out that change is needed, and the comments may provide an area to research. Finally, the teams must agree upon which topics to research next, who should be involved, which insights would deserve new design improvements, and how to prioritize future backlog items to be implemented. If any step is halted, then the improvements may not be achieved. These steps can be owned by different departments or roles, yet there should be alignment among them. 

Different roles may have different viewpoints. For example, the UX team might want the UX improvements to be implemented, while the sales team wants new features to be added, and the dev team wants to support a new technology. All viewpoints can be valid and should be balanced. If the UX KPIs are reasonably high, then the team can agree to reduce focus on the UX improvements and instead to put focus on other areas of the product. However, if the UX KPIs are quite low, the product team should agree on what UX KPI target to hit and in which timeframe, and then support activities to reach this goal. 

There needs to exist trust amongst the team members where all have the same goal to improve the product, even though different team members may have different ideas about how to get there. Team members should be open minded to see the perspectives of others. If the product team jointly sets the UX goal, they jointly control how quickly and how much effort they are willing to invest. 

After the improvements are implemented and released, further survey responses will assess these improvements. Note the scores may take time to rise because fixing a single issue will not cause a drastic change in scores. It is only through consistent and sustained improvements where the users will start rating the products higher.

\section{Outlook}
A best-case scenario for gathering UX feedback with the help of surveys might look like this: UX KPIs are well established and accepted in a company, from the board-level to the product teams. This is because UX KPIs and the improvement of UX KPIs have proven to correlate with other business relevant KPIs, such as contract renewals, or monthly active users, and product teams notice that their joint efforts of investing in UX projects lead to increased UX KPIs. Executives understand that UX KPIs often require more than just a few quarters to improve. Product teams understand that improving the UX is a joint responsibility, involving all disciplines. Improving the UX can concern various aspects: It may range from investing in improving the stability, reliability, or performance of a product, to harmonizing workflows across different areas of a product or to integrating innovative features, to name just a few examples. Product teams use the open text feedback to identify areas of improvement, and apply different methods to fill knowledge gaps, qualitative user research being one of them. They also iterate constantly with users when designing new workflows or features. Users continuously provide feedback because they notice that the company takes the feedback seriously and invests in projects to improve the UX.  

We believe there are a few factors that increase the chance to get to such a best-case scenario, many of them summarizing factors named above:

\begin{itemize}
\item Ensure high-level executive buy-in for the UX KPIs early on.
\item Include the UX KPIs in the general company KPIs and KPI setting process.
\item Involve stakeholders on all levels early on in setting up the KPIs, the process, and your assets.
\item Establish a network of champions across different products that help to implement and evangelize your UX KPI project.
\item Make the data from the UX surveys transparent and accessible for everyone with the help of dashboards and reports.
\item Provide support and enablement for teams to make use of the data.
\item Analyze the relationship between UX KPIs and business relevant KPIs and make them transparent.
\item Analyze the impact of UX improvement initiatives on UX KPIs and celebrate if they were successful.
\item Continuously gather feedback from your stakeholders and improve your initiative. 
\end{itemize}

Every company is different and may require different approaches and focus areas. But the end-to-end process described in this article has worked for us so far and comprises generic factors that should apply and work for software companies in general. Gathering UX feedback from users is a key prerequisite to improving the UX, and eventually, further business relevant KPIs.

Moving forward, the recent advancements in the field of generative AI, especially large language models, provide enormous opportunities to derive further insights and automate the interpretation of user feedback in the form of comments. Creating summaries of comments automatically, for example, greatly facilitates the processing of text data. Additionally, the categorization of comments as well as the analysis of sentiments can help to generate further insights with relatively low effort. In combination with the other UX metrics, the automated analysis of user comments provide a comprehensive overview of how users perceive the UX of a product. Leveraging these insights facilitate the effective planning of product improvements.

\section*{Acknowledgements}
The authors gratefully acknowledge the support of Viola Stiebritz on the graphical design of some figures in this publication. We also thank Michael Ameling for useful comments on the text.



\bibliographystyle{mnras}
\bibliography{LinkedIn_bib} 








\label{lastpage}
\end{document}